\documentclass[aps,showkeys]{revtex4}
\usepackage{epsfig}

\begin{document}

\title{Simulations on the elastic response of amorphous and nanocomposite 
carbon}

\author{Maria Fyta\footnote{Present address: Department of Physics, 
Harvard University, 17 Oxford Str., Cambridge, 02138 MA, USA.} and Pantelis C. Kelires}
\address{Physics Department, University of Crete, P.O. Box 2208, 
710 03, Heraklion, Crete, Greece}

\begin{abstract}
Theoretical calculations of the elastic response of carbon composites 
and amorphous carbon are reported. The studied composites consist of 
crystalline nanoinclusions, either spherical diamonds or carbon nanotubes,
embedded in amorphous carbon matrices. The elastic constants of the 
composites were calculated and found to systematically increase as the 
density increases. The elastic recovery under hydrostatic pressure
for all structures was also investigated and was found to be
significantly high for both nanocomposite and amorphous carbon, but 
decreases as the material becomes more dilute.

\end{abstract}


\maketitle
%\end{frontmatter}

\section{Introduction}

Novel carbon materials are studied intensively due to 
their enormous potential utility. Nanocomposite films are 
materials formed by various carbon crystallites 
embedded in an amorphous carbon ($a$-C) matrix. 
Interest on these materials is growing since their properties, 
both mechanical and electronic are expected to tailor and improve 
the properties of the single-phase $a$-C system. 
Depending on the density and the specific structure of the 
crystalline inclusion, the properties of the two parts of the
system can intermingle leading to materials with diverse 
and interesting properties.

The nanoinclusions can range from $sp^3$ conformations to entirely
$sp^2$ ones, which in the absence of a matrix may interact {\it via}
van der Waals (VDW) forces. Carbon composites have been studied 
theoretically \cite{nc@aC,CNT@aC,bulknD@aC}, while there is also insight 
from experimental work. Diamond composites formed by a nanometer sized
diamond surrounded by $a$-C are potential candidates 
for superhard materials with applications
in electromechanical devices such as MEMS/NEMS or as coatings.
Such composites have recently been grown \cite{Lifshitz} in
both hydrogenated and pure $a$-C matrices. 
On the other hand, carbon nanotubes (CNT) \cite{cnt}, which are fully
$sp^2$ forms of carbon have attracted many 
experimental and theoretical studies as they have inspired 
interesting advances in science. Although their
production methods are continuously improving, in many
cases not only pure CNT but also other crystalline
as well as amorphous carbon structures are formed.
Thereby, either intentionally or not, CNTs can be found in $a$-C, 
forming films with potentially improved mechanical, thermal, and
electronic properties for technological applications.

In this work, we study some mechanical properties of carbon nanocomposites.
Recently, we investigated how diamond composites respond to the
application of strain, in the regime where fracture occurs, and we
extracted their strength \cite{bulknD@aC}. Here, instead, we focus on
the elastic regime, i.e., the region where the stress is still directly 
proportional to the strain applied. The elastic constants and moduli
and the elastic recovery are the representative quantities to unravel
the behavior of the two nanomaterials within this regime. 

\section{Numerical set-up}

Computer simulations are used to model various $a$-C networks
and carbon composites. The latter contain either
a spherical diamond or a CNT, in order to capture the effects of a 
fully $sp^3$ and a fully $sp^2$ inclusion on the properties
of the composite. Monte Carlo calculations within the 
empirical potential approach are carried out. These are effective
methods for the analysis of structural and mechanical properties 
of composite systems. The atomic interactions are modeled by 
the Tersoff potential \cite{tersoff}. 
{In the case of nanotube composites 
(CNT/$a$-C), a typical 12-6 Lennard-Jones potential was also 
added to effectively model the interaction of the CNT atoms 
and those of the surrounding matrix \cite{CNT@aC}. The parameters used for 
this potential ($\epsilon$=2.964 meV, $\sigma$=3.407 \AA) have been 
previously used to describe the bulk properties of solid $C_{60}$ 
and multiwall nanotubes \cite{ljparam}.} 

This scheme is capable of describing 
large systems, while it is also well tested and provides a 
fairly good description of the structure and energetics of a 
wide range of crystalline carbon and $a$-C phases 
\cite{nc@aC,kelires94,Donadio}. 
In particular, the Tersoff potential has been fitted and
well tested with respect to reproducing the elastic constants, defect 
energies and phonon frequencies of diamond and graphite. Work
by Mathioudakis and Kelires \cite{MathiouJNCS} showed that the
potential describes well the variation and softening of elastic
moduli of amorphous semiconductors as a function of temperature.
In addition, it gives valuable 
information on the elastic properties and stress in various 
a-C forms \cite{kelires94,kelires00}.
These issues are closely related and valuable to 
the present work. However, the potential does not describe $\pi-\pi$ 
interactions properly, due to the absence of dihedral-angle forces,
giving rise to excessive $sp^{2}-sp^{3}$ mixed hybrids. 
The absence of repulsion between non-bonded $\pi$ 
orbitals \cite{Stephan,Kaukonen,Jager} leads to some overestimation of 
the density for a given sp$^3$ content \cite{kelires00,Mathiou}. This mostly 
occurs at intermediate sp$^3$ contents, while low-sp$^3$ (evaporated a-C) and 
high-sp$^3$ (tetrahedral ta-C) networks are properly described in this respect. 
For ta-C (sp$^3$ fraction about 80\%) the overestimation
is $\sim$ 3-4\%. This shortcoming is not expected to seriously affect
the energetics and elastic moduli of the nanocomposites, because these
are determined primarily by the $\sigma$ bond character of the network, 
which is very well treated by the potential.

\begin{figure}
\begin{center}
\epsfig{file=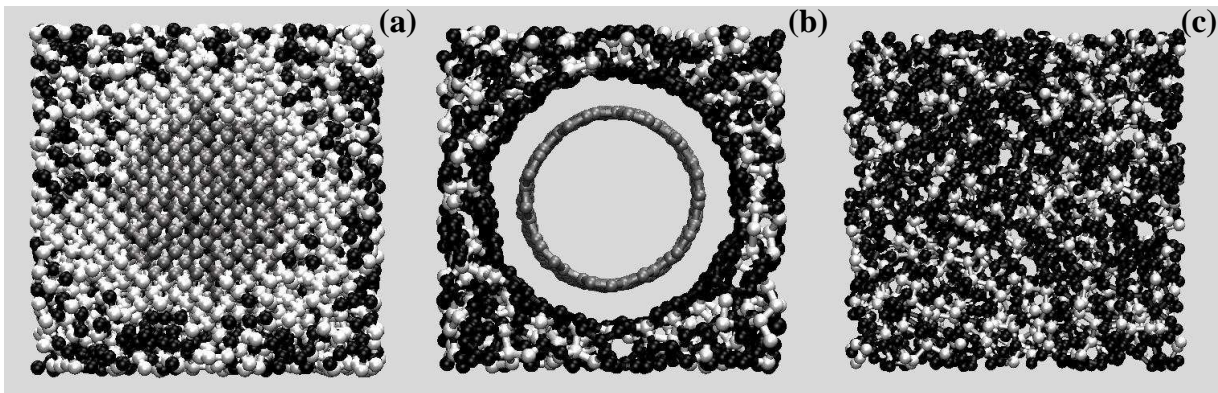,width=0.9\textwidth}
\caption{Representative ball \& stick models of the structures
investigated here: (a) $n$D/$a$-C with an $sp^2$ fraction about 15\%, 
(b) CNT/$a$-C  and (c) single-phase $a$-C, both with $sp^2$ fractions
about 30 \%. The $sp^3$ atoms are shown in white, $sp^2$ in black, while
in (a) and (b) the grey atoms are those of the crystalline inclusions.
In (a) some of the matrix atoms are not shown, in order to reveal
the diamond core.}
\label{fig1}
\end{center}
\end{figure}

All structures were generated by quenching from the melt, a simulational 
method that is widely used to generate amorphous
networks with a variety of different methods, ranging from
first principles calculations or tight binding schemes to more empirical
methods as the one presented in this paper.
This procedure, produces generic structures that cannot be directly 
associated with the nonequilibrium as-deposited structures, but can be 
definitely associated with and can be used to study the equilibrium 
ground state of carbon composite films, i.e., with structures sufficiently
relaxed under moderate thermal annealing. 
The generated structures depend on applied conditions, such
as the quenching rate. This issue has been examined thourougly in 
previous studies \cite{nc@aC,kelires93}. We use the optimized rates 
obtained in these studies. The structures also depend on the applied
pressure (variable volume simulation) or the chosen initial density
(constant volume). We are able to produce a wide range of
$sp^3$ fractions by appropriately choosing the density/pressure of
the simulation.

In the case of nanocomposites, the crystalline 
atoms of the inclusion were kept frozen throughout the quenching process.
After quenching, which produces amorphization of the surrounding matrix, the 
cells are thoroughly relaxed with respect to atomic positions and density.
 Relaxations are particularly important at the interface region, where the
 crystallites mainly adjust to the host environment. 
One way of checking the stability of nanocrystals, 
is to subject them to thermal annealing. A stable structure 
should be sustained in the amorphous matrix, while an unstable structure 
should shrink in favor of the host.

Details have been given elsewhere \cite{nc@aC,CNT@aC},
where it was also shown that diamond crystallites are stable
(under conditions of thermodynamic equilibrium) in dense $a$-C 
matrices \cite{nc@aC}, while nanotubes are stable (or slightly 
metastable) in low density matrices \cite{CNT@aC}.
Here, we are interested only on these stable composites, which could
be used for practical purposes.

Spherical diamonds or single-wall CNTs of different sizes are embedded
into amorphous matrices of various densities. Cubic cells with periodic 
boundary conditions, containing 500-5000 atoms, were used and the diameters 
of the crystalline inclusions were allowed to vary. 
All the steps in the generation of the 
structures was done with the Monte Carlo
method within the Tersoff empirical potential approach.

The nanodiamond diameters range from 0.5 to 2 nm, while for nanotubes 
they lie within 0.8-2.7 nm. 
 For all structures, the initial distance between 
the inclusions due to the periodicity of the cells is in the
range 0.9 - 1.1 nm. After the generation of the matrix and the relaxation,
this distance will vary as some $a$-C atoms at the 
crystalline-amorphous interface are expected to crystallize
(dense composites) or atoms of the inclusions may amorphisize
(more dilute composites).  Representative models of all 
different structures, both composites and single-phase $a$-C, 
that have been studied here are shown in Fig.\ref{fig1}.
Simulations on multi-wall nanotube composites were also
carried out. These will not be extensively analyzed, but will
serve as a comparison to single-wall nanotube composites.
All properties are calculated at a low temperature. 
The $sp^3$ ($sp^2$) compositions
of the structures given in the following analysis
refer to the amorphous matrices, while
the densities refer to the whole structure, unless
otherwise stated. The amount of $sp$ atoms for all cases studied
here is negligible and the threefold ($sp^2$) atoms can be directly
estimated from the fourfold ($sp^3$) atoms.

\section{Elastic Constants}

The elastic constants and moduli of diamond and nanotube composites are 
studied using standard techniques \cite{elastConst}
at the lowest-energy configurations. These supplement previous
results on their bulk moduli \cite{CNT@aC,bulknD@aC} and are compared
to the a-C values \cite{kelires94}. Strain 
is applied to the cubic cells and the elastic energy directly
proportional to the elastic constants is obtained. The 
strain is small and leads to small deviations from cubic symmetry, 
allowing us to refer to cubic elastic moduli.
A volume conserving orthorombic strain was applied to calculate
the shear modulus equal to $c_{11}-c_{12}$, while the $c_{44}$ elastic
constant is given by shearing the cell with a monoclinic strain.
The equilibrium bulk modulus $B$ was extracted through a fit to the 
energy-versus-volume curve using the
Birch-Murnaghan isothermal equation of state \cite{murnachan}.
The Young's modulus is calculated from the bulk modulus
through the relation $Y=3B (1-2\nu)$, with $\nu=C_{12}/C_{11}+C_{12}$
being the Poisson ratio.
This is defined as the ratio of the contraction normal 
to the applied strain divided by the extension in the 
direction of the applied strain. According to the theory of isotropic 
elasticity, its values should be in the range from -1 to 0.5. 
We show here that this ratio for 
diamond and nanotube composites is between 0.1 and 0.3.
The calculated $B$ for bulk diamond and for
a free standing CNT of $\sim$ 1 nm, using the
Tersoff potential, is 443 and $\sim$ 200 GPa, respectively. These
values are in very good agreement with experimental results
for the former, and with {\it ab initio} simulations for the 
latter \cite{bulkCNT}. 

\begin{table}
\begin{center}
\caption{Moduli (in GPa) of diamond composites with a 1.7 nm diamond 
inclusion in a dense matrix with 80\% $sp^3$ content.}
\begin{tabular}{|c|c|c|c|c|c|c|c|}
\hline
$\rho$(g cm$^{-3})$&B&$c_{11}$&$c_{12}$&$c_{44}$&shear&$\nu$&Y\\ \hline
2.6&215&493&76&210 & 208&0.13&473\\
2.7&241&566&78&240 & 244&0.12&547\\ 
2.9&268&636&84&289 & 276&0.12&616\\ 
3.1&307&735&93&341 & 321&0.11&714\\
3.3&344&817&107&366 & 355&0.12&792\\ 
3.4&359&856&110&387 & 373&0.11&831\\ \hline
\end{tabular}
\label{table1}
\end{center}
\end{table}

We begin with calculations related to diamond composites.
The trends followed by the bulk moduli are similar to those
given by more accurate tight-binding simulations \cite{bulknD@aC}.
As a general result, all the elastic
moduli show an increase as the structure becomes more
dense or the inclusion larger. For example, a nanodiamond with
$\sim$ 1.7 nm in diameter embedded
in a dense matrix (with 80\% $sp^3$ content) has a bulk modulus close 
to 360 GPa and a Young's modulus of 830 GPa. The corresponding values for 
a nanodiamond with a diameter of 2.1 nm in a similar matrix rise up to 370 
and 865 GPa, respectively. The shear modulus of these two composites 
increases from 373 to 388 GPa.
In order to elucidate the trend of increasing moduli with increasing
density, the calculated values for a composite with a
1.7 nm diamond inclusion are presented in Table \ref{table1}.
These clearly reveal the hardening
of the composite as the $a$-C matrix becomes more dense and the
behavior is similar to the case of pure $a$-C \cite{kelires94}.
The effect of the diamond core on the properties of a composite 
should almost be the same for all densities (especially for the 
stable dense nanocomposites) and the variation of the elastic
constants with density should mainly be determined by the matrix, which 
becomes more incompressible as it gets denser. 
However, for low densities (high $sp^2$ fraction) the moduli
of the composites are lower than those of single-phase $a$-C
networks with the same density. For high densities, the increase over
the $a$-C values is $\sim$ 8\% for a 1.7 nm diamond inclusion
and 13\% for a 2.1 nm nanodiamond.
Hence, in dense composites, both the inclusion and the surrounding
matrix are quite incompressible and the whole structure is quite rigid,
while in the opposite limit not only the matrix becomes more dilute,
but the inclusion becomes more deformed and contributes to the
softening of the whole structure. These softening mechanisms
are mainly related to $sp^2$ sites, which contribute less to the
rigidity of the material \cite{kelires00}. Therefore, composites with low 
density (high $sp^2$ content) matrices have elastic 
constants that are lower than those of single-phase $a$-C. On the other
hand, the Poisson ratio shows no significant fluctuations indicative
of the fact that the enhancement in both $c_{11}$ and $c_{12}$
is similar and contraction versus extension is almost the same
for all $n$D/$a$-C structures.
Also, the shear value is smaller than the bulk modulus
for the same structure, verifying what is known for
most materials, i.e., they resist a change in volume (determined
by the bulk modulus) more than they resist a change in shape.
This is also the case for nanotube composites.

A first look at these CNT/$a$-C structures reveals that their
elastic moduli are in general smaller than those of $n$D/$a$-C. 
In our opinion, two
are the main reasons behind this trend. Diamonds are
stable in high density, thus more rigid matrices, but
CNTs can be found mainly in less dense (more soft) 
matrices \cite{CNT@aC}.
A second important factor in CNT/$a$-C is related to the empty 
regions of appreciable volume inside and directly outside the
nanotube were no atoms occur. As shown in Ref.\ \cite{CNT@aC}, no
covalent bonding between CNT atoms and the matrix exists. 
Instead, the VDW forces drive a reconstruction of the surrounding
medium, producing a graphitic wall at a distance of 3.4 \AA from the
CNT. See panel (b) of Fig. 1. Thus, the resulting open space (voids),
together with the relative high $sp^2$ ratio  lower the
moduli of such materials, rendering them softer.
Again, similar to the bulk modulus trends, all elastic parameters
of the dense composites are enhanced.
Some of the results are summarized in Table \ref{table2},
where Y$_{\parallel}$ and Y$_{\perp}$ are the contributions 
in the Young's modulus in directions along and perpendicular
 to the tube axis \cite{calc}. It is clear that for 
the same nanotube diameter all moduli 
increase as the density of the whole material increases. This 
strenghtening is again necessitated by the more dense
(more rigid) $a$-C matrix.
Here, the Poisson ratio is slightly higher than for
diamond composites and its fluctuations are more significant.
This is again a signature of the higher anisotropy that
the nanotube composites show.
The transverse contraction over the longitudinal extension 
when strain is applied to CNT/$a$-C is higher than
for $n$D/$a$-C.
Comparison of similar, in terms of density, CNT/$a$-C structures
reveals again an enhancement of the moduli as the CNT 
diameter increases. 
For example, a composite with a nanotube 1.2 nm in diameter 
has a shear modulus of 79 GPa and a Young's modulus of 273 GPa, 
while these values rise to 131 and 416 GPa for a 1.7 nm CNT.
Both composites have a density of 2 g cm$^{-3}$.
We note that different nanotube chiralities did not affect any of the
results, thereby the nanotubes are rather anonymous and we
refer only to their diameter.

At this point, a comment regarding the differences between
single and multi-wall nanotube composites has to be made.
The trends followed by the latter composites are similar to the former
ones, since they
also become more rigid for higher densities and larger nanotube 
diameters. Nevertheless, there is an additional feature 
that plays a crucial
role in the rigidity of multi-wall nanotube composites. This is
related to the number of walls that form a multi-wall nanotube. 
As the number of walls increases, while the total nanotube diameter
remains constant, the elastic moduli increase because of the
repulsive forces between the CNT walls at short distances.

\section{Rule of mixtures}

Thus far, the contribution of the a-C matrix and its $sp^2$
component in the determination of the elastic
parameters was examined. In this section, we seek to find out 
in what extent the properties of each constituent 
are mapped on the properties of the whole structure.
Nanocomposites are a mixture of two or more different phases. It has been
proposed that in such configurations the elastic moduli of the whole 
structure is proportional to the sum of the moduli of each of the 
constituents with their volume fractions serving as the weights in 
the summation \cite{Tabor}. This is known as the {\it rule of mixtures}.
We now investigate whether this rule applies here by comparing 
the moduli presented in the preceding section to the ones given
through this rule.
A diamond nanocomposite with a crystalline volume fraction of 33\% and
a total density $\sim$ 3.3 g cm$^{-3}$ ($sp^{2}$ component
about 25\%) has a modulus of 346 GPa. The rule
of mixtures for this composite leads to a value of 354 GPa. Similar are
the results for all dense structures. As the density
decreases, the deviation of the calculated moduli from those
given through the rule of mixtures becomes larger.
For a diamond composite with a total $sp^{3}$ fraction
of 50 \% this difference is close to 40 GPa and gets
larger with a decreasing $sp^{3}$ fraction. These features indicate
that diamond inclusions are not strongly affected by their presence in
high density $a$-C matrices, while this is not the case for
dilute $a$-C matrices in which the nanodiamonds are unstable 
and quite distorted, thus influenced by the embedding effect.
\begin{table}
\begin{center}
\caption{Elastic constants (in GPa) of composites with nanotube inclusions.}
\begin{tabular}{|c|c|c|c|c|c|c|c|c|c|c|c|}
\hline
D(nm)&$\rho$(g cm$^{-3}$)&$C_{11}$&$C_{12}$&$C_{44}$&B&Y&$\nu$&Y$_{\parallel}$&Y$_{\perp}$\\ \hline
1.2&2.2&337&121&79&193&273&0.26&476 &290 \\ 
1.2&2.4&418&104&108&209&377&0.20&601 &364 \\
1.6&2.0&290&78&71&149&257&0.21&478  &231 \\ 
1.6&2.1&384&81&94&182&356&0.17&564&306\\ 
2.1&1.9&295&79&97&151&262&0.21&490&258 \\ 
2.7&1.7&251&72&82&132&219&0.22&404& 216\\ \hline
\end{tabular}
\label{table2}
\end{center}
\end{table}

On the other hand, the rule of mixture
does not hold for composites with nanotube inclusions.
This is a direct evidence that CNTs
are strongly influenced by their presence in the mixture
and show a stability which is in average lower than that 
for diamond inclusions.
The axial $Y_{\parallel}$ and transversal $Y_{\perp}$ Young's
 moduli of a composite with a SWNT inclusion of a 1.65 nm diameter 
are 564 GPa and 306 GPa, respectively.
The values obtained by the rule are $Y_{\parallel}$ = 297 GPa and
 $Y_{\perp}$ = 323 GPa. In general, the axial moduli compared to 
the transverse part
deviate more from the values calculated through the rule of mixtures
and the anisotropy of the elastic moduli of free standing nanotubes is
enhanced by their embedding into the matrix. In
the axial direction the embedding effect is stronger and leads to a 
strengthening along this direction, while a slight softening is evident
 along transverse directions. 
The matrix is an isotropic medium and the anisotropy is 
induced by the CNT. It is stronger for the axial part of the 
Young's modulus
due to the infinite length of the CNT in this direction,
along which the CNT actually dominates over the matrix.
This happens regardless of the density.
In transverse directions, the matrix contribution is stronger 
as it is leads to a softening along these directions.

\section{Elastic Recovery}

Finally, we discuss the elastic recovery during a 
compression-decompression
cycle of both composite and amorphous structures. 
Experimentally, this is often performed by atomic force microscopy or
standard nanoindentation and nanoscratch experiments.
Measurements for $a$-C films show that these recover
highly elastic ($\sim$ 90\%) \cite{recovExpr}, but no
variation with the $sp^2$ fraction is known. In our knowledge there are
no such experiments for the composites studied here and 
our results serve as predictions for the elastic recovery
of these materials.
Computationally, by applying hydrostatic pressure on these composite
structures their contraction can be measured. A permanent deformation 
signifies a decrease of the volume expressed as a fraction of its
initial value. A perfectly elastic material would have an elastic 
recovery equal to one, while for a perfect plastic one it would be
zero. CNT ropes, as an example, have been found to have high elastic 
recovery even under high-deformation 
conditions (over 100 GPa) \cite{CNTelastRecov}.

In order to reveal the elastic response of the composites,
compression of all these stuctures up to 100 GPa took place,
followed by full decompression.
The behavior is directly dependent on the stability of the composites, 
i.e., the more stable the structure the higher its recovery.
We mainly focus on the stable composites, since these
are the ones that are of technological interest. It is now more 
efficient to refer to the total $sp^3$ component for each case
in order to promote comparison with the composite stuctures.
Diamond composites are stable for an $sp^3$ fraction over 
60-70\%, while for nanotube composites this value is in the
range of 20-30\%. The elastic recovery of single-phase $a$-C is determined
by its density or hybridization. It is almost perfectly elastic
in the highly tetrahedral regime ($ta$-C).
The recovery is not perfect as the $sp^3$ fraction declines, but
it is again high in good aggreement to the experimental
results mentioned previously. An almost fully 
$sp^2$-bonded $a$-C network 
regains its initial volume at about 95 \%. 
For a much higher compression (up to 500 GPa) the dense networks
again recover almost perfectly, but for the dilute ones (fully $sp^2$)
the recovery drops to about 80 \%. Similarly, in diamond composites,
the more stable these are the higher their recovery. It is again
almost perfect (even for compression up to 500 GPa), 
but as their density
lowers the recovery drops to $\sim$ 90\%. Taking into account
the high incompressibility of the diamond crystallites the
decrease can only be assigned to the matrix and mainly
to its threefold atoms, which are more easily deformed.

The response of nanotube composites is more interesting. Although,
they recover their initial volume over 90\%, there are essential
configurational modifications after decompression. The CNT retains
its shape, but a few covalent bridge bonds ($\sim$ 1.6 \AA) between 
the CNT and the matrix have been formed. During decompression
the volume is changed (up to $\sim$ 10\% depending on density)
denoting that 
the system has been trapped into a local minimum. 
The bridge bonds, which start to form
at a pressure close to 70 GPa, modify the structure and do
not allow volume changes during decompression. In the case of 
unstable nanotube composites the volume change after 
decompression is again low, but inspection of the nanotube reveals 
the strong effect inflicted by the pressure.
The tubular character is preserved, but the cross section is far 
from being circular, although no buckling similar to free-standing CNTs 
undercompression was evident \cite{buckl}. The matrix and especially its 
part that immediately surrounds the inclusion is not seriously affected
and no bridge bonds were formed.

\begin{figure}
\begin{center}
\epsfig{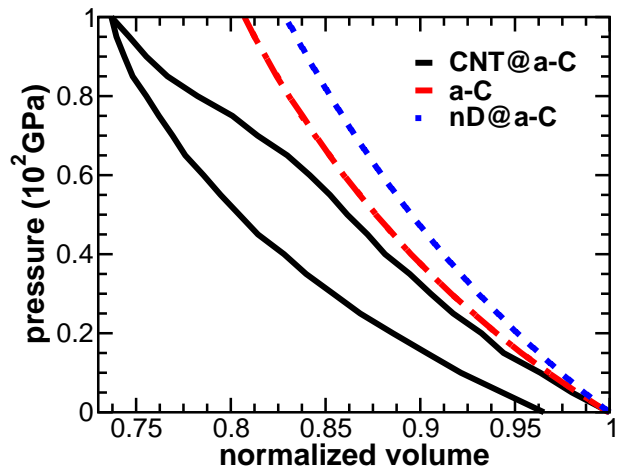}
\caption{Compression - decompression curves for $n$D/$a$-C 
(dotted/blue), CNT/$a$-C (solid/black) and 
single-phase $a$-C (dashed/red).}
\label{fig2}
\end{center}
\end{figure}

A sample of all cases studied is summarized in Fig.\ref{fig2}. 
The elastic recovery for an $a$-C network with $\sim$ 40\% 
threefold atoms, as well as that for diamond and nanotube 
composites are shown. 
The composites shown were chosen according
to their stability. In the case of
$a$-C, we show a network that serves as an upper bound
(in terms of the $sp^2$ fraction)
to the perfectly elastic $a$-C samples. More threefold atoms
lead to deviations from this behavior as mentioned previously. 
The diameters of the inclusions and the
$sp^2$ fractions are 1.7 nm and 10\% for the diamond composite,
and 1.2 nm and $\sim$ 60\% for the nanotube composite. The number
of threefold sites in the case of the diamond composite is
small, but the elastic recovery is similar for large
components, as long as the composite retains its stability. 
The compression curves for $a$-C and $n$D/$a$-C coincide to those
for decompression, indicating the full elastic recovery
of these structures. The curves related to
CNT/$a$-C show a hysterisis representing the change in the final volume
after decompression.
This behavior is related to the structural modification in the final 
configuration mentioned previously.
Nanotube composites can be more easily compressed as shown in the
figure. At the high pressure of 100 GPa, their volume can be reduced up 
to 25\% from its initial value. The reduction for single-phase $a$-C is 18\%,
while it is even less for diamond composites. 
For the latter, the $sp^2$ component is much smaller compared to the
CNT/$a$-C and $a$-C networks, but even for a larger component
the compressibility is small. (A $n$D/$a$-C network with 
73\% of its atoms being $sp^2$ is compressed up to 77\% at 100 GPa.
This value is again higher than it is for $a$-C and
CNT/$a$-C.) Our results indicate that the compressibility from CNT/$a$-C 
towards $a$-C and $n$D/$a$-C is decreasing. The crucial quantity
for the degree of recovery is the $sp^2$ content. As long as it remains
high, there is no perfect elastic recovery. Here, we only 
correlated this threefold component to the elastic recovery
of the composites and $a$-C. A next step would include a more
extensive study of the size of the nanoinclusion.

\section{Conclusions}

A comparative study of the elastic constants of diamond and nanotube
composites was presented. A systematic increase in these properties
was found as the structures become more dense. The amorphous
matrix and mainly the $sp^3$ component contribute the most
to this enhancement. Study of the elastic response
of all composite structures and single-phase $a$-C revealed a 
high elastic recovery. It is almost perfect for the stable
composites and the amorphous networks, but decreases
with increasing $sp^2$ component.
Nanotube composites are more easily deformed, followed by the
pure $a$-C networks and the diamond composites. The higher
the $sp^2$ fraction the higher the compression, which eventually
leads to a lower elastic recovery.

\acknowledgments

We thank G. Hadjisavvas for useful discussions and S. Kassavetis for 
helpful insights into his $a$-C related experimental work.
The work was supported by a grant from the EU and the 
Ministry of National Education and Religious Affairs of Greece through the 
action ``E$\Pi$EAEK'' (programme ``$\Pi\Upsilon\Theta$A$\Gamma$OPA$\Sigma$''.)


\begin{thebibliography}{}
\bibitem{calc} The empty regions were subtracted from the volume 
of the composite. The volume of the CNT was assumed equal
to 2$\pi$R$\delta$R, with $\delta$R close to the graphite 
interplanar distance, a value widely used in the literature;
 R is the radius of the CNT after full relaxation of the cell.
\end{thebibliography}
\end{document}